\documentclass[10pt,twocolumn,letterpaper]{article}

\usepackage{cvpr}              

\usepackage[accsupp]{axessibility}  

\definecolor{cvprblue}{rgb}{0.21,0.49,0.74}
\usepackage[pagebackref,breaklinks,colorlinks,allcolors=cvprblue]{hyperref}

\def\paperID{} 
\def\confName{CVPR}
\def\confYear{2026}

\title{Adversarial Agents: Black-Box Evasion Attacks with Reinforcement Learning}

\author{
    Kyle Domico\textsuperscript{1},
    Jean-Charles Noirot Ferrand\textsuperscript{1},
    Ryan Sheatsley\textsuperscript{1}, \\
    Eric Pauley\textsuperscript{2},
    Josiah Hanna\textsuperscript{1},
    Patrick McDaniel\textsuperscript{1} \\[6pt] 
    \textsuperscript{1}University of Wisconsin-Madison \quad
    \textsuperscript{2}Virginia Tech\\
    {\tt\small \{domico, jcnf, sheatsley, jphanna, mcdaniel\}@cs.wisc.edu, pauley@cs.vt.edu}
}

\begin{document}
\maketitle
\begin{abstract}


Attacks on machine learning models have been extensively studied through stateless optimization. In this paper, we demonstrate how a reinforcement learning (RL) agent can learn a new class of attack algorithms that generate adversarial samples. Unlike traditional adversarial machine learning (AML) methods that craft adversarial samples independently, our RL-based approach retains and exploits past attack experience to improve the effectiveness and efficiency of future attacks. We formulate adversarial sample generation as a Markov Decision Process and evaluate RL's ability to (a) learn effective and efficient attack strategies and (b) compete with state-of-the-art AML. On two image classification benchmarks, our agent increases attack success rate by up to 13.2\% and decreases the average number of victim model queries per attack by up to 16.9\% from the start to the end of training. In a head-to-head comparison with state-of-the-art image attacks, our approach enables an adversary to generate adversarial samples with 17\% more success on unseen inputs post-training. From a security perspective, this work demonstrates a powerful new attack vector that uses RL to train agents that attack ML models efficiently and at scale.
\end{abstract}

\section{Introduction}\label{introduction}
The advancement of AI has led to an explosion of applications that rely on the decision-making and generation capabilities of ML models. Such models have revolutionized online assistants, commerce, content generation, cybersecurity, entertainment and gaming, to name just a few.  At the same time, adversaries wishing to manipulate these applications have developed algorithms that produce inputs that are intended to fool the model, e.g., bypass a content filter. Adversarial machine learning (AML) studies algorithms for generating adversarial samples targeted at victim models~\cite{papernot_limitations_2016, carlini_towards_2017, madry_towards_2017}. Extant AML uses gradient-based optimization to minimize the distortion applied to an input such that the result is misclassified by the victim ML model~\cite{chen_zoo_2017,chen_hopskipjumpattack_2020, croce_reliable_2020, vedaldi_square_2020}.  However, the technical community's understanding of model defenses and adversarial capabilities is at best limited.

\begin{figure}
    \centering
    \includegraphics[width=\linewidth]{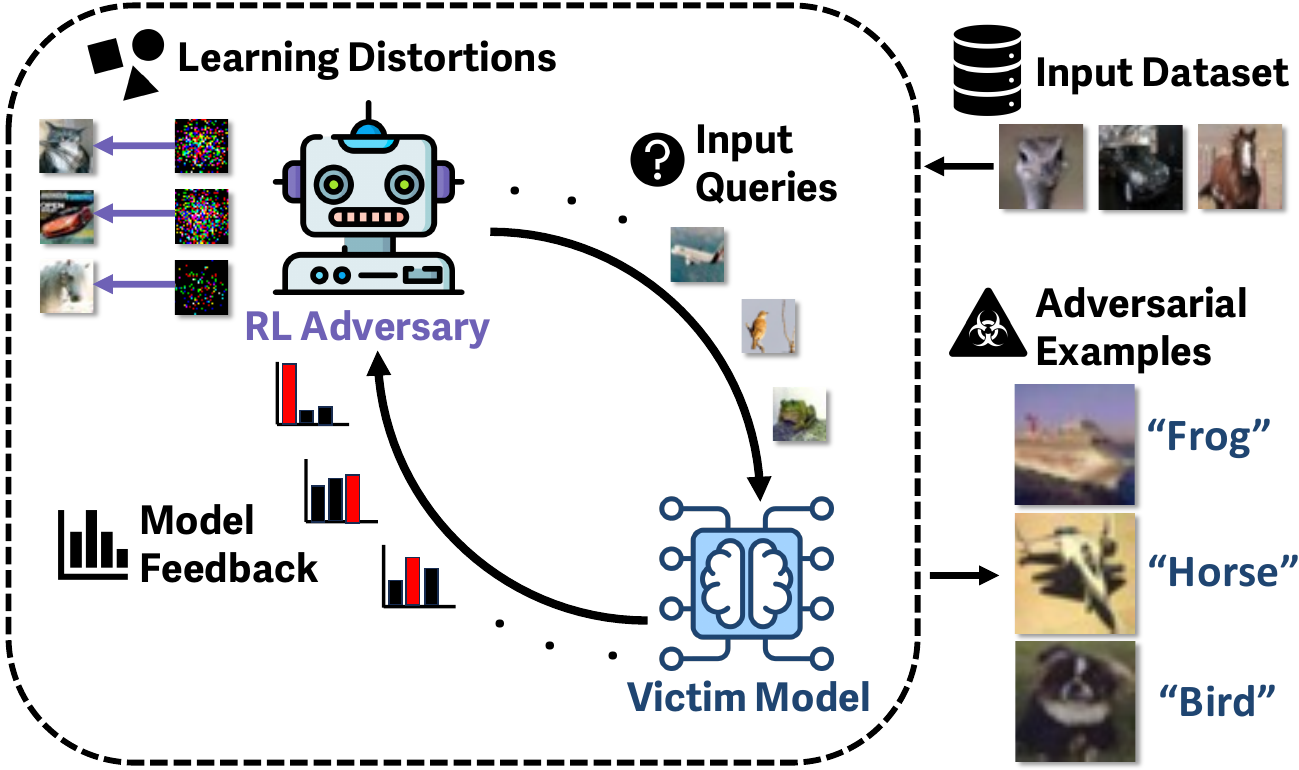}
    \caption{Overview of the reinforcement learning attack with CIFAR-10. The adversary interacts with the victim model by querying images and receiving feedback, iteratively generating adversarial samples.}
    \label{fig:grid}
\end{figure}

Most existing adversarial algorithms are stateless, treating each target input as a one-off optimization problem. This presents a gap in AML, as the adversary is unable to learn and improve its strategy from past experience. For example, consider an advanced persistent threat (APT) agent that uses reinforcement learning (RL) to evade facial recognition systems with restricted query access (i.e., black-box access). This sustained interaction allows the APT agent to learn a \textit{generalizable} evasion strategy that successfully attacks multiple inputs, extending beyond the limitations of single-sample attacks. Such learned persistence and efficiency would represent a new capability for adversaries attacking machine learning systems.

We posit that modeling the adversary as a reinforcement learning (RL) agent will enable attack strategies that become increasingly more efficient and effective over time, a capability not currently available in existing work. In this paper, we introduce and evaluate an attack based on RL primitives to generate adversarial samples. Modeled as an RL agent, the adversary learns which perturbations are the most effective at fooling the model given an arbitrary input. Once the agent is trained, the adversary then uses the learned policy to generate adversarial samples. Thus, an adversarial agent can attack a model without costly gradient optimization.

We demonstrate that adversarial sample generation can be modeled as a Markov Decision Process (MDP). The MDP formulation allows us to readily use RL in such a way that encapsulates the underlying semantics of the attack: inputs and victim model outputs as states, perturbations as actions, and differences in adversarial objectives as rewards. We introduce two distinct MDPs, RL Max Loss and RL Min Norm, each designed to achieve a different adversarial objective. For both MDPs, we use the Proximal Policy Optimization (PPO) algorithm~\cite{schulman_proximal_2017} to train an agent, which is then used in a policy evaluation setting to craft adversarial samples. The attacks are demonstrated on two benchmark image classification tasks against \texttt{ResNet50}~\cite{he_deep_2015}, \texttt{VGG-16}~\cite{simonyan_very_2015}, and \texttt{ViT-B/16}~\cite{dosovitskiy_image_2021} victim models. Within this framework, we conduct a multi-step evaluation of the agent, assessing its performance across (a) learning capabilities, (b) hyperparameter tuning, and (c) accuracy relative to a standard black-box attack algorithm.

First, we evaluate whether an RL agent can learn strategies that improve black-box evasion attacks. In other words, can RL improve the effectiveness and efficiency of adversarial samples over training? Indeed, both RL Max Loss and RL Min Norm attacks increase discounted return over training, validating RL's ability to learn the task. Throughout training, the rate at which adversarial samples are produced increases by up to 13.2\%, while the amount of interaction with the victim model decreases by up to 16.9\%. These results empirically show that agents grow stronger with training by producing more adversarial samples efficiently. We illustrate adversarial agent training with RL on CIFAR-10 samples in \autoref{fig:grid}.

Next, we analyze how MDP hyperparameters in both RL Max Loss and RL Min Norm affect adversarial samples. The $\epsilon$ parameter controls the amount of distortion the RL Max Loss agent is allowed to have on a given input. The $c$ parameter controls the degree to which the RL Min Norm agent gets rewarded for minimizing distortion over reducing victim model confidence. While training the agent increases the performance of both attacks, we show that trained agent performance depends on the choice of $\epsilon$ and $c$. For our experiments, we choose $\epsilon$ and $c$ according to this sensitivity analysis that balances adversarial objectives. An adversary using these attacks must consider them before attacking to meet their desired goal.

Lastly, we evaluate how the trained agent's ability to craft adversarial samples generalizes to unseen data and how it performs relative to traditional query-based black-box attack algorithms~\cite{andriushchenko_square_2020, chen_hopskipjumpattack_2020, ilyas_prior_2019}. On an unseen dataset, the trained agent's attack success rate, average queries, and average distortion remain consistent with the distribution observed during the training of adversarial samples. In a black-box comparison with the optimization-based baselines, we compare the trained agent's performance on unseen data to show that using RL to improve black-box evasion attacks enables the adversary to generate up to 17\% more adversarial samples with 31\% fewer queries. These results highlight the efficacy of an adversary learning from past attack experience through RL over existing methods.
\section{Background and Related Work}\label{sec:background}
Evasion attacks in AML aim to craft adversarial samples: inputs, typically with human-imperceptible modifications, that cause a model to make an incorrect prediction~\cite{szegedy_intriguing_2014, goodfellow_explaining_2015}. Crafting these samples involves a trade-off between two primary objectives: (a) maximizing the victim model's prediction error for a given distortion budget (Max Loss), or (b) finding the minimum possible distortion required to cause a misclassification (Min Norm). Formally, given an input x, its true label y, and the victim model classifier Z, these objectives are defined as:
\begin{equation}
    \begin{aligned}
        & \text{\textbf{Max Loss: }} \arg\max_{\delta} L(Z(x+\delta), y) \text{ s.t. } \|\delta\|_p \leq \epsilon \\
        & \text{\textbf{Min Norm: }} \arg\min_{\delta} \|\delta\|_p \text{ s.t. } \arg\max Z(x+\delta) \neq y
    \end{aligned}
\end{equation}

where $\delta$ is the adversarial perturbation, $L$ is a loss function (e.g., cross-entropy), and $\epsilon$ is the maximum distortion budget for a given $\ell_p$-norm. The method for solving these optimizations depends on the adversary's knowledge, or threat model. In a white-box setting, full access to the victim model's architecture and parameters enables efficient, gradient-based methods like Projected Gradient Descent (PGD) for Max Loss attacks~\cite{madry_towards_2017} and the Carlini \& Wagner (C\&W) attack for Min Norm problems~\cite{carlini_towards_2017}.

In the more challenging black-box setting, the adversary has only query access to the model's outputs. These outputs can be hard-label class predictions (decision-based) or class probabilities (score-based). For black-box attacks, minimizing the number of queries becomes a third critical objective alongside maximizing model loss and minimizing distortion. A prominent score-based black-box method used in our evaluation is Square Attack~\cite{andriushchenko_square_2020}, which solves the max loss problem by iteratively applying random, localized square-shaped updates to the input.

Prior applications of RL to adversarial tasks have focused on the traditional AML setting of optimizing over inputs independently~\cite{tsingenopoulos_autoattacker_2019, sarkar_robustness_2023, hore_deep_2025}. We argue that these works understate the true agentic capability of RL in this domain: learning attack strategies that improve with experience and generalize to new data. By drawing a parallel between the objective of AML optimizations and the RL process, our work introduces a generalizable framework built on this paradigm of learning from past attacks. To demonstrate the practicality of this approach, we conduct a utility analysis on data outside the RL agent's training environment and present a direct comparison of attack success and query-efficiency against established black-box attack methods.

\section{Learning Adversarial Policies}\label{methodology}
RL enables the adversary to learn from the success and failure of crafting adversarial samples. Black-box settings involve a feedback loop between the adversary and victim model, allowing us to construct the MDP necessary to use RL. The rest of the section will lay out (1) the episodic setting with the MDP, and (2) the attack procedure.

\subsection{MDP Formulation}\label{attack_mdp}
In the adversary's MDP, each episode begins with a randomly sampled clean input as part of the initial state. During an episode, the adversary applies perturbations to a select set of features and is rewarded in part by the decrease in victim model confidence on the true label of the selected image. Episodes terminate after the victim model is fooled, meaning that an adversarial sample has been crafted (or a threshold of steps is reached, and the process has been deemed to fail). Here, we show an episode with a random agent and randomly sampled input from CIFAR-10~\cite{krizhevsky_learning_nodate}:

\centerline{\includegraphics[width=\linewidth]{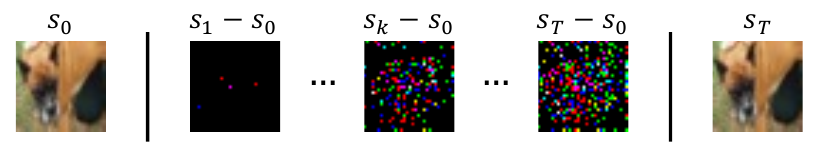}}

Concretely, the processes begin with an input at the start state ($s_0$). At each timestep ($s_1 \dots s_k$), the agent applies (human imperceptible) incremental perturbations to the input that lowers the victim model confidence. The episode terminates in state ($s_T$) when the model changes its classification to the adversary selected label, or a maximum of steps is reached.  Next, we show how we can model this process as an MDP.

\noindent\textbf{States and Actions.}
The state space of the attack MDP consists of information available to the adversary. In the black-box setting, this includes the victim model outputs on a given sample $Z(\cdot)$ (i.e., oracle). We define the state representation $s_t$ of the attack at a given timestep $t$ as:
\begin{equation}\label{state_space}
	s_t = (x_t, y, Z(x_t))
\end{equation}
where $x_t \in \mathbb{R}^n$ is the resulting input after $t$ steps of the environment starting at input sample $x_0$, $y$ the ground-truth label of $x_0$, and $Z(x_t)$ the victim model outputs on $x_t$. This information allows the agent to learn effective perturbations at different states of the process. The actions in the environment represent a distortion to be applied to the current input $x_t$. Learning fine-grained distortions on every feature of the input becomes difficult on high-dimensional data (e.g., images, text, or network data). To create a more tractable learning problem, we simplify the action space by having the agent select a small subset of $N$ features to perturb. This dimensionality reduction is critical for query efficiency in black-box settings, as searching the full high-dimensional input space is often infeasible~\cite{ilyas_prior_2019}. We define the action $a_t$ at a given timestep $t$ as a set of $N$ feature-perturbation pairs:
\begin{equation}\label{action_space}
    a_t = \left\{(i_{1}, \delta_{1}), (i_{2}, \delta_{2}),\dots, (i_{N}, \delta_{N})\right\}
\end{equation}
where $i_{1}, ..., i_{N} \in \{1,n\}$ are the indices of the $N$ input features selected for modification. Each selected feature $i_j$ paired with a distortion $\delta_j$, with each having magnitude $|\delta_j|\leq \theta$ for $j\in\{1,...,N\}$. This allows us to balance the trade-off between many and few distorted features with large or little distortion through $N$ and $\theta$ for effective and efficient policy learning.

\noindent\textbf{Reward and Transition Functions.}\label{rewards_transitions}
As highlighted above, we categorize AML objectives into two classes: Max Loss and Min Norm. Our goal is to shape reward and transition functions so that RL optimizes these objectives. To quantify the victim model's confidence of the true label, we use:
\begin{equation}
    f(x,y) = \log([Z(x)]_y)
\end{equation}
where $f$ measures the log probability of the victim model's confidence of $x$ belonging to class $y$ (i.e., negative cross-entropy loss). To model one-step differences in victim model confidence and distortion, let us define:
\begin{equation}
    \begin{aligned}
        & \Delta_{t+1} f=f(x_t,y)-f(x_{t+1},y) \quad \\
        & \Delta_{t+1}\delta=\|x_t-x_0\|_2-\|x_{t+1}-x_0\|_2
    \end{aligned}
\end{equation}
where $x_{t+1}$ is the input at the next state $s_{t+1}$ and we use the $\ell_2$-norm to quantify distortion. For simplicity, we define the function $\phi:\mathbb{R}^n \times \mathcal{A}\to\mathbb{R}^n$ that takes an input sample $x_t\in\mathbb{R}^n$ and action $a_t\in \mathcal{A}$ and returns the input with specified distortions from the action. The following will describe two versions of an RL attack corresponding to the adversaries' goals.

\noindent\textbf{RL Max Loss.} This setting rewards actions that reduce the victim models confidence. The transition should project the distortion onto a $\epsilon$ distortion budget centered at the starting input $x_0$ and keep the distortion if the victim model's confidence is reduced. Let us define $x_t^{a_t}=\text{Proj}_\epsilon[\phi(x_t,a_t)-x_0]+ x_0$ as the candidate input compliant with the $\epsilon$ budget constraint after action $a_t$ is applied. Then, the next input $x_{t+1}$ at next state $s_{t+1}$ can be defined with resultant reward $R(s_t,a_t)$ as 

\begin{equation}
x_{t+1} =
\begin{cases}
x_t^{a_t} & \textbf{if } f(x_t,y) - f(x_t^{a_t},y) > 0 \\
x_t       & \textbf{otherwise}
\end{cases}
\end{equation}
\begin{equation}
    R(s_t,a_t) = \Delta_{t+1}f
\end{equation}

where the input at the next state and the reward follow the distortion success in reducing the victim model confidence with the proposed action. It is important to note that a transition involves exactly one query to the victim model to collect outputs $Z(x_t)$. We denote RL Max Loss transition and reward as $P_{\text{Max Loss}}$ and $R_{\text{Max Loss}}$, respectively.

\noindent\textbf{RL Min Norm.} This setting should give rewards to actions that not only reduce victim model confidence but minimize the distortion in doing so. Like RL Max Loss, the state of the next input should result from applying an action that reduces victim model confidence with minimal distortion. Let us define $x_t^{a_t}=\phi(x_t,a_t)$ as the candidate input. The next input $x_{t+1}$ at the next state $s_{t+1}$ can be defined with resultant reward $R(s_t,a_t)$ as

\begin{equation}
x_{t+1} =
\begin{cases}
  x_t^{a_t} & \textbf{if } \begin{aligned}[t]
                         & c \cdot \big[\|x_t-x_0\|_2 - \|x_t^{a_t}-x_0\|_2\big] \\
                         & + \big[f(x_t,y) - f(x_t^{a_t},y)\big] > 0
                       \end{aligned} \\
  \\ 
  x_t & \textbf{otherwise}
\end{cases}
\end{equation}
\begin{equation}
    R(s_t,a_t) = c\cdot\Delta_{t+1}f + \Delta_{t+1}\delta
\end{equation}

where $c$ weights the importance of minimizing distortion over reducing the victim model confidence. We experiment with different values of $c$ in the evaluation to determine a value that balances the effectiveness and efficiency of distortions. Like RL Max Loss, a transition involves one query and the distortion proposed by the action is kept if the change in victim model loss and distortion is positive and rewarded accordingly. We denote RL Min Norm transition and reward as $P_{\text{Min Norm}}$ and $R_{\text{Min Norm}}$, respectively.

\begin{figure}[t]
    \centering
    \includegraphics[width=\linewidth]{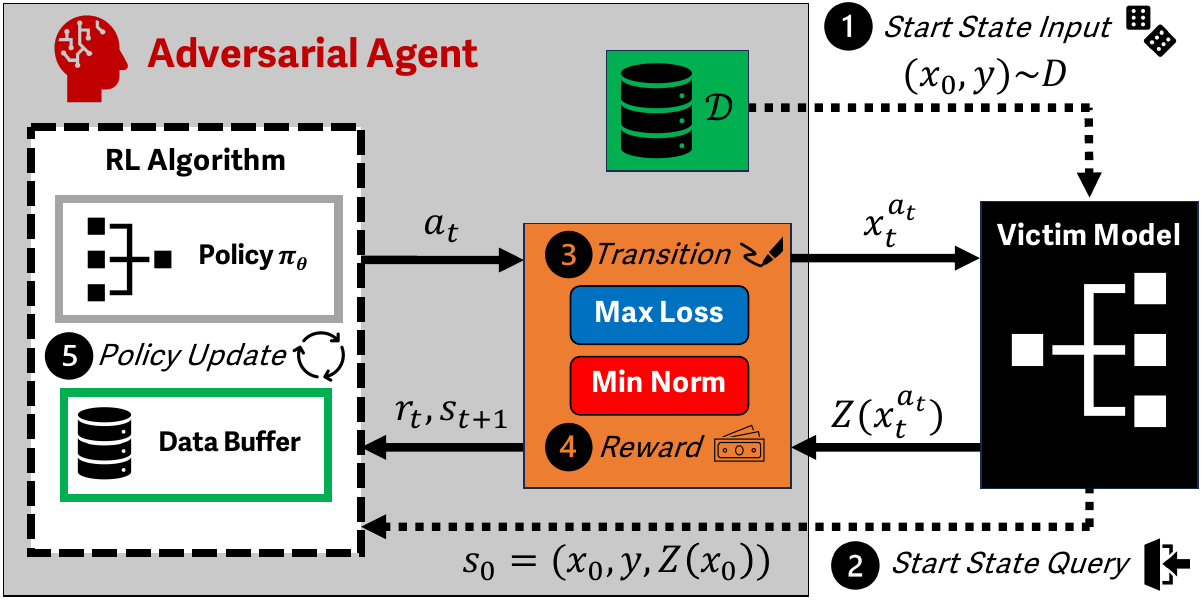}
    \caption{Training the adversarial agent: (1) a randomly sampled clean input and label $(x_0,y)$ from training dataset $\mathcal{D}$, (2) the start state $s_0=(x_0,y,Z(x_0))$ is initialized with the initial query of $x_0$, (3,4) the transition and reward function with respect to RL Max Loss and RL Min Norm mediate interaction with the victim model, (5) the policy $\pi_{\theta}$ is updated according to the RL algorithm.}
    \label{fig:methodology}
\end{figure}

\subsection{The Reinforcement Learning Attack}
We leverage these MDPs to launch the RL attack. The attack comes in two forms: (a) attacking while training the agent\footnote{Note that in practice, an attacker would train the agent by accessing the model in a subtle way to avoid detection (e.g., by querying the model slowly over time or using known model training data to create a surrogate model to train the agent,~\cite{papernot_practical_2017}).  Thereafter, the adversary could use the trained agent to attack the model without restraint.}
and (b) attacking with a fixed policy. In \autoref{fig:methodology}, we detail the attack during the training stage. First, a randomly sampled input and label $(x_0,y)$ from training dataset $\mathcal{D}$ is used to query the victim model to initialize the start state $s_0=(x_0, y, Z(x_0))$. The $(s,a,r,s^\prime)$ interaction with the victim model proceeds following policy $\pi_{\theta}$ with respect to the attack type RL Max Loss or RL Min Norm. The agent stores interactions and updates the policy according to the specific RL algorithm (on-policy or off-policy). When the episode terminates, the process restarts at start state sampling. After training the agent, the adversary can use the trained adversarial agent to perform the attack in a traditional RL policy evaluation setting. The next section evaluates how well this approach trains adversarial agents that learn better attack strategies.
\section{Evaluation}\label{eval}
We evaluate the adversarial algorithm by asking the following research questions:
\begin{enumerate}
    \itemsep0em 
    \item \textit{Do adversarial agents learn more effective and efficient attacks during training?}
    \item \textit{How do key environment hyperparameters ($\epsilon$ and $c$) influence the trade-off between attack effectiveness and efficiency?}
    \item \textit{How well do trained adversarial agents generalize to unseen data, and how does it compare to traditional black-box methods?}
\end{enumerate}

\subsection{Experimental Setup}\label{experimental_setup}

\begin{figure*}[t]
    \centering
    \includegraphics[width=0.99\linewidth]{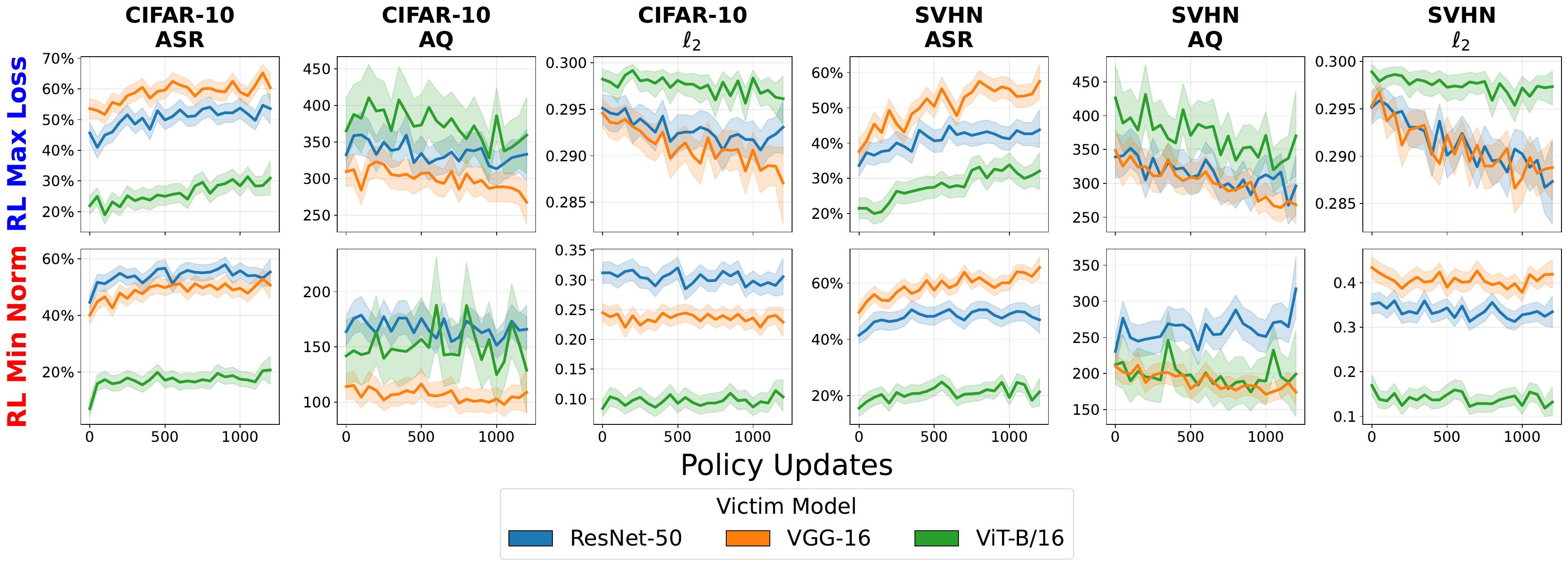}
    \caption{\textcolor{blue}{RL Max Loss} and \textcolor{red}{RL Min Norm} attack training: attack success rate (\textbf{ASR}), average queries on successful attacks (\textbf{AQ}), and average $\ell_2$-norm distortion on successful attacks ($\ell_2$) with respect to policy updates for 3 trials per attack with a 95\% confidence interval on CIFAR-10 and SVHN datasets.}
    \label{fig:learning}
\end{figure*}

\begin{table}[t]
\centering
\small
\begin{tabular}{l|c|c}
\toprule
\textbf{Model Architecture} & \textbf{CIFAR-10} & \textbf{SVHN}\\
\midrule
\texttt{ResNet-50}~\cite{he_deep_2015} & 97.1\% & 96.4\% \\
\texttt{VGG-16}~\cite{simonyan_very_2015} & 94.7\% & 95.2\% \\
\texttt{ViT-B/16}~\cite{dosovitskiy_image_2021} & 97.9\% & 97.3\% \\
\bottomrule
\end{tabular}
\caption{Test Classification Accuracy of Victim Models: comparison of victim models across the standard test split of image classification datasets.}
\label{tab:model_accuracy}
\end{table}

\noindent\textbf{Datasets.}
We use two image classification datasets in our experiments: CIFAR-10~\cite{krizhevsky_learning_nodate} and Street View House Numbers (SVHN)~\cite{noauthor_reading_nodate}. CIFAR-10 is an object recognition dataset and is widely used in AML literature. It is made up of 60 000 training and 10 000 testing samples. Each sample has 3 072 features encoding pixel values. For black-box experiments, we use the 10 000 test samples and create two partitions: 5 000 samples for $\mathcal{D}$ in attack training, and 5 000 samples as a hidden dataset $\mathcal{D}^\prime$ for policy evaluation. SVHN is a digit recognition dataset collected from Google Street View images by cropping out digit sequences from house number plates. The data set consists of 73,257 samples for training and 26,032 for testing. Like CIFAR-10, each sample has 3 072 features encoding pixel values and we create two partitions of the test samples: 13 016 for $\mathcal{D}$ in attack training, and 13 016 as a hidden dataset $\mathcal{D}^\prime$.

\noindent\textbf{Model Architectures.}
We experiment with three different victim models $Z$: \texttt{ResNet-50}~\cite{he_deep_2015}, \texttt{VGG-16}~\cite{simonyan_very_2015}, and \texttt{ViT-B/16}~\cite{dosovitskiy_image_2021}. For our experiments, we use models pre-trained on the ImageNet-1K dataset~\cite{deng_imagenet_2009} and fine-tune them on the individual target datasets. The resulting test accuracies are reported in \autoref{tab:model_accuracy}. For RL training, we use the PPO implementation in StableBaselines3~\cite{raffin_stable-baselines3_2021}. The adversarial agent's policy consists of a \texttt{EfficientNet}~\cite{tan_efficientnet_2020} feature extractor followed by a fully connected feedforward neural network that outputs the parameters of the action distribution. The PPO algorithm also uses a separated value function network which we model as a feedforward network that shares the policy's \texttt{EfficientNet} feature extractor. For training adversarial agents, we perform our experiments on NVIDIA A100 GPUs with 40 GB of VRAM.


\noindent\textbf{Action Hyperparameters.}
Our methodology defines actions as selecting $N$ features of the input with associated features $\delta_i$ for $i\in\{1,..,N\}$. For our evaluation, we fix $N=5$ and $\theta=0.05$, as preliminary experiments showed these values offered a good balance between attack success and action complexity (see supplementary material for selection details). We focus on evaluating the agents ability to learn and generalize adversarial strategies.

\noindent\textbf{Metrics.}
Adversarial samples in all threat models are benchmarked on distortion and victim model misclassification. As such, we use the $\ell_2$ distortion and attack success rate (ASR) (i.e., $\arg \max_i Z(x_T)^{(i)} \neq y$) at terminal states $s_T$. The exception appears with black-box threat models, where the number of victim model queries per adversarial sample are also considered. Thus, we determine the average queries to victim models on episodes that produce successful attacks (AQ). These metrics encapsulate the strength of a black-box adversary and will be used to show their changes as the agent is trained.

\subsection{Training Adversarial Agents}\label{learning}
In this section, we ask: \textit{do adversarial agents learn more effective and efficient attacks during training?} In other words, is the agent able to learn a policy with greater attack success, and does the adversary get stronger with more successful adversarial samples in fewer queries?  To answer this, we perform an evaluation in two parts: (1) analyzing policy learning and attack effectiveness over training, and (2) analyzing attack efficiency over training. We use ASR, AQ, and $\ell_2$ distortion of adversarial samples as metrics. In \autoref{fig:learning}, we train RL Max Loss and RL Min Norm agents in training datasets $\mathcal{D}$ for 1200 PPO policy updates with 3 random seeds each and plot the mean and 95\% confidence interval over attack metrics. We fix $\epsilon =0.3$ for all RL Max Loss experiments, $c=10^{-2}$ for RL Min Norm on CIFAR-10, and $c=10^{-3}$ for RL Min Norm on SVHN. \autoref{fig:learning} organizes the plots with the dataset and metrics on the columns and the RL attack type on the rows. We observe that the agent is learning (a) a more effective policy if attack success increases, (b) a more query-efficient policy if successful episode queries decrease, and (c) a more distortion-efficient policy if $\ell_2$ distortion decreases.

\noindent\textbf{Effective Learning.}
Our goal is to show that the ASR for each attack increases over training to determine if the adversarial agents learn more effective attacks. Here, we focus on the \textbf{ASR} columns of \autoref{fig:learning}. Note that the ASR scales differ across RL attack type and dataset. We observe that the RL Max Loss attack consistently increases ASR over training across all models and datasets with averages of 8.7\% increase on \texttt{ViT-B/16}, 9.3\% increase on \texttt{ResNet-50}, and 13.2\% increase on \texttt{VGG-16}. Similarly, we observe that the RL Min Norm attack increases the attack success with averages of 5.1\% increase on \texttt{ViT-B/16}, 5.6\% increase on \texttt{ResNet-50}, and 9.7\% increase on \texttt{VGG-16}. We note that the difference in performance between RL attack types is because the RL Max Loss attack enforces the $\ell_2$ budget constraints in the state transition while RL Min Norm must learn it through the reward feedback. Nevertheless, the results demonstrate RL's ability to learn effective strategies for generating adversarial samples.

\noindent\textbf{Efficient Learning.}
With our black-box evaluation metrics for adversarial samples, we aim to show that adversarial samples produced throughout training become more efficient. Here, we focus on the AQ and $\ell_2$ distortion columns of \autoref{fig:learning}. Note that the scale of successful queries and distortion differ across RL attack type and dataset. We observe that RL Max Loss improves query-efficiency significantly throughout training with up to 11.3\% and 16.9\% decrease in average queries on successful episodes on CIFAR-10 and SVHN tasks, respectively. In contrast, the RL Min Norm attack improves query efficiency only on \texttt{VGG-16} models by an average of 6.6\%. However, these steep changes in query-efficiency for RL Max Loss tells us that the adversarial agents learn a query-efficient attack strategy.

In the $\ell_2$ distortion plots, we observe the same trends for each attack on both CIFAR-10 and SVHN tasks. RL Max Loss consistently stays below its $\epsilon=0.3$ budget constraint, confirming that the agent efficiently utilizes the allowable distortion to increase attack effectiveness and query-efficiency. The decrease in distortion observed in all models and datasets is influenced by the agent requiring fewer queries over training. Contrarily, the $\ell_2$ distortion RL Min Norm requires for attacks varies widely on the dataset and victim model. This demonstrates that the agent, while achieving marginal learning gains, struggles to efficiently minimize distortion and queries simultaneously when those constraints are only guided through the reward function, rather than being explicitly enforced in the state space like in RL Max Loss.

\begin{figure}
    \centering
    \includegraphics[width=\linewidth]{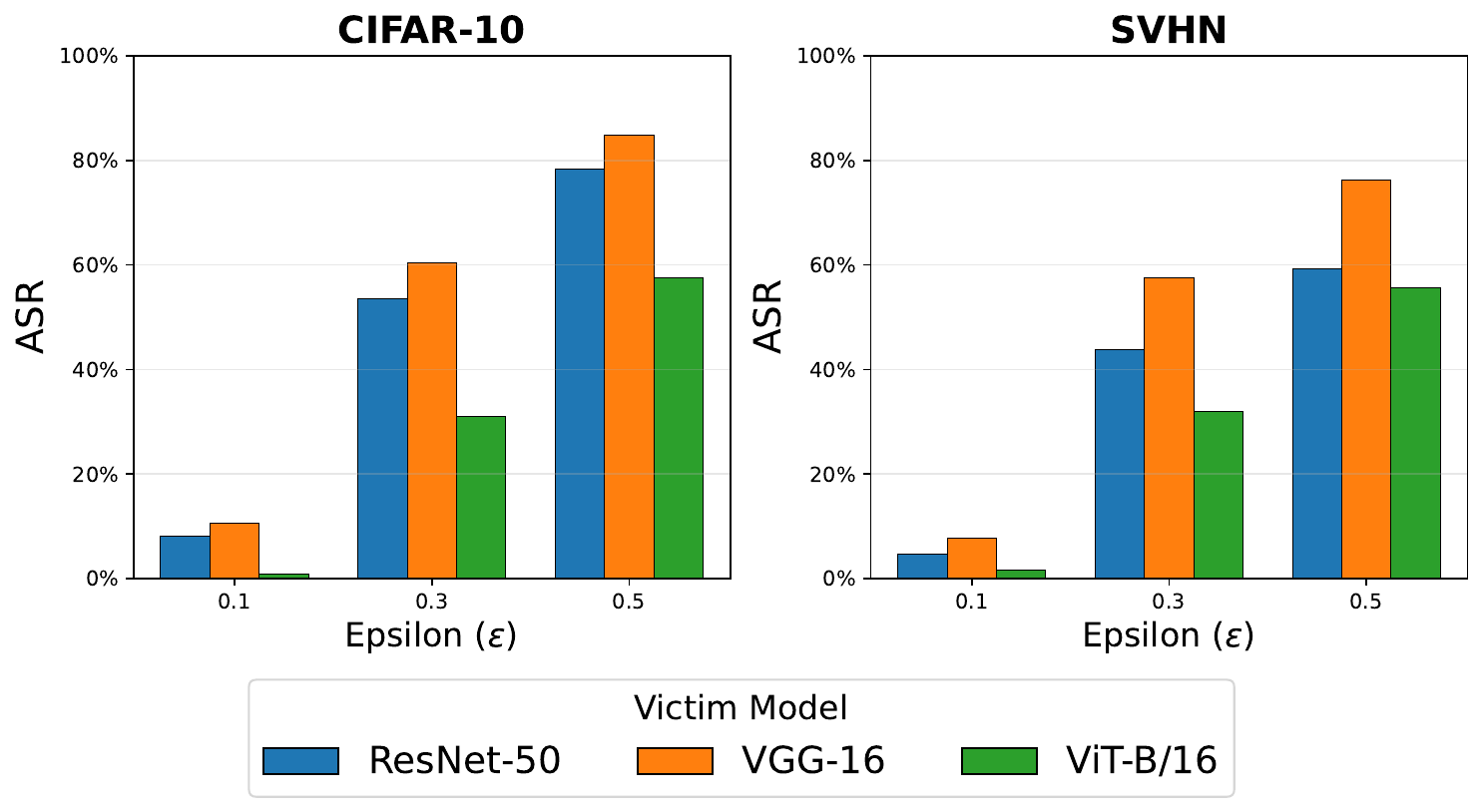}
    \caption{\textcolor{blue}{RL Max Loss} Hyperparameter Sensitivity: attack success rate (ASR) versus Epsilon ($\epsilon$) for trained agents averaged over 3 random seeds.}
    \label{fig:epsilon_sensitivity}
\end{figure}

\begin{figure}
    \centering
    \includegraphics[width=\linewidth]{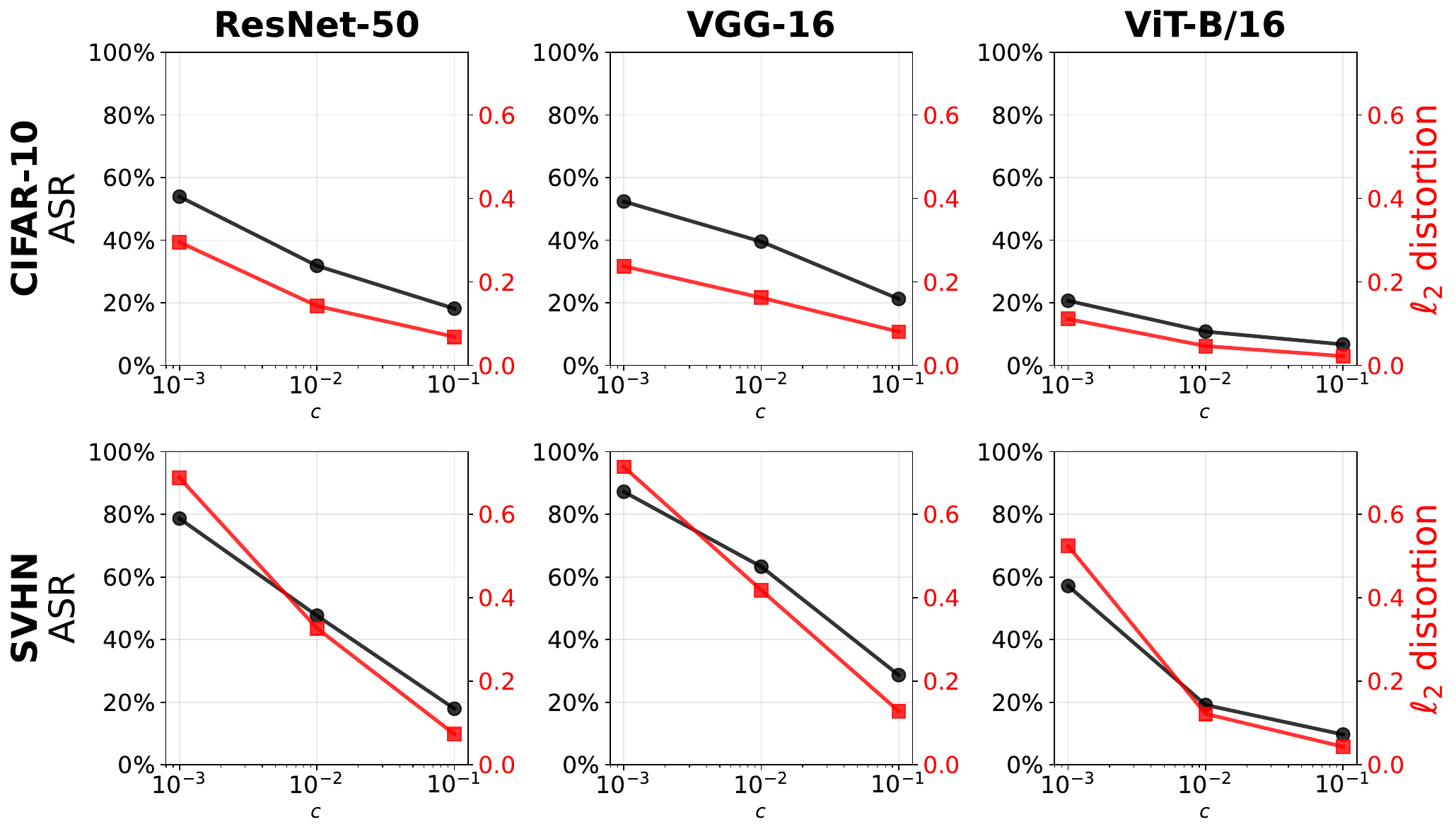}
    \caption{\textcolor{red}{RL Min Norm} Hyperparameter Sensitivity: attack success rate (ASR) and $\ell_2$ distortion vs. $c$ for trained agents averaged over 3 random seeds.}
    \label{fig:c_sensitivity}
\end{figure}

\begin{table*}[h]
\centering
\small 
\setlength{\tabcolsep}{3.5pt}
\begin{tabular}{l|ccc|ccc|ccc|ccc|ccc|ccc}
\toprule
& \multicolumn{9}{c|}{\textbf{CIFAR-10}} & \multicolumn{9}{c}{\textbf{SVHN}} \\
\cmidrule(lr){2-10} \cmidrule(lr){11-19}
& \multicolumn{3}{c|}{\texttt{ResNet-50}} & \multicolumn{3}{c|}{\texttt{VGG-16}} & \multicolumn{3}{c|}{\texttt{ViT-B/16}} & \multicolumn{3}{c|}{\texttt{ResNet-50}} & \multicolumn{3}{c|}{\texttt{VGG-16}} & \multicolumn{3}{c}{\texttt{ViT-B/16}} \\
\cmidrule(lr){2-4} \cmidrule(lr){5-7} \cmidrule(lr){8-10} \cmidrule(lr){11-13} \cmidrule(lr){14-16} \cmidrule(lr){17-19}
\textbf{Attack \& Dataset} & \textbf{ASR} & \textbf{AQ} & $\ell_2$ & \textbf{ASR} & \textbf{AQ} & $\ell_2$ & \textbf{ASR} & \textbf{AQ} & $\ell_2$ & \textbf{ASR} & \textbf{AQ} & $\ell_2$ & \textbf{ASR} & \textbf{AQ} & $\ell_2$ & \textbf{ASR} & \textbf{AQ} & $\ell_2$ \\
\midrule
\textcolor{blue}{RL Max Loss} ($\mathcal{D}$) & 0.55 & 295 & 0.29 & 0.64 & 265 & 0.29 & 0.30 & 321 & 0.30 & 0.43 & 287 & 0.29 & 0.59 & 278 & 0.29 & 0.32 & 336 & 0.30 \\
\textcolor{blue}{RL Max Loss} ($\mathcal{D}'$) & 0.59 & 315 & \textbf{0.30} & \textbf{0.64} & 259 & 0.30 & 0.28 & 332 & 0.30 & 0.45 & 285 & \textbf{0.30} & 0.60 & 255 & \textbf{0.30} & \textbf{0.32} & 344 & 0.30 \\
\midrule
\textcolor{red}{RL Min Norm} ($\mathcal{D}$) & 0.58 & 168 & 0.32 & 0.53 & 111 & 0.25 & 0.19 & 155 & 0.11 & 0.49 & 265 & 0.34 & 0.65 & 191 & 0.41 & 0.20 & 187 & 0.15 \\
\textcolor{red}{RL Min Norm} ($\mathcal{D}'$) & \textbf{0.62} & \textbf{155} & 0.35 & 0.55 & \textbf{103} & \textbf{0.29} & 0.17 & \textbf{148} & \textbf{0.12} & \textbf{0.53} & \textbf{249} & 0.36 & \textbf{0.65} & \textbf{186} & 0.42 & 0.18 & \textbf{192} & \textbf{0.18} \\
\midrule
Square~\cite{andriushchenko_square_2020} ($\mathcal{D}^\prime$) & 0.53 & 335 & \textbf{0.30} & 0.61 & 350 & 0.30 & \textbf{0.31} & 344 & 0.30 & 0.48 & 263 & \textbf{0.30} & 0.48 & 273 & \textbf{0.30} & 0.29 & 262 & 0.30 \\
HSJA~\cite{chen_hopskipjumpattack_2020} ($\mathcal{D}^\prime$) & 0.31 & 681 & \textbf{0.30} & 0.36 & 637 & 0.30 & 0.13 & 904 & 0.30 & 0.30 & 705 & \textbf{0.30} & 0.38 & 639 & \textbf{0.30} & 0.12 & 923 & 0.30 \\
Bandits~\cite{ilyas_prior_2019} ($\mathcal{D}^\prime$) & 0.44 & 635 & \textbf{0.30} & 0.47 & 662 & 0.30 & 0.08 & 516 & 0.30 & 0.39 & 556 & \textbf{0.30} & 0.46 & 537 & \textbf{0.30} & 0.14 & 589 & 0.30 \\
\bottomrule
\end{tabular}
\caption{Performance comparison of RL-based attacks and black-box attack baselines on the CIFAR-10 and SVHN datasets across different victim models: \texttt{ResNet-50}, \texttt{VGG-16}, and \texttt{ViT-B/16}. We report the attack success rate (\textbf{ASR}), average queries on successful attacks (\textbf{AQ}), and average $\ell_2$-norm distortion on successful attacks ($\ell_2$) on training set $\mathcal{D}$ and testing set $\mathcal{D}^\prime$.}
\label{tab:attack_comparison}
\end{table*}

\subsection{Sensitivity Analysis}\label{sensitivity}
The two types of RL attacks share the same state and action representations, but differ in reward and transition functions. Recall that the $\epsilon$ parameter in RL Max Loss controls the maximum distortion allowable, and $c$ in RL Min Norm controls the weight of the reward given to reducing victim model confidence over minimizing distortion. Here, we ask: \textit{how do key environment hyperparameters ($\epsilon$ and $c$) influence the trade-off between attack effectiveness and efficiency?}

This analysis is broken down into two parts: (1) analyzing the effect $\epsilon$ has on the adversarial samples crafted by RL Max Loss, and (2) analyzing the effect $c$ has on the adversarial samples crafted by RL Min Norm. We use the ASR and $\ell_2$-norm distortion on adversarial samples crafted after training to examine how changing $\epsilon$ and $c$ affects adversarial capabilities. We plot the performances of trained agents in policy evaluation settings for RL Max Loss and RL Min Norm in \autoref{fig:epsilon_sensitivity} and \autoref{fig:c_sensitivity}, respectively.

\noindent\textbf{Analyzing the effect of $\epsilon$.}
With our setup of the RL Max Loss attack, the agent is rewarded by $\Delta_{t+1}f$ at each timestep and all distortions stay within $\epsilon$ $\ell_2$-norm distortion budget through $R_{\text{Max Loss}}$ and $P_{\text{Max Loss}}$. We run the RL Max Loss attack for 1200 PPO policy updates with $\epsilon \in \{0.1,0.3, 0.5\}$ and tested on the training images from $\mathcal{D}$. We evaluate the policies trained from 3 random seeds and plot the average ASR with respect to $\epsilon$ on each victim model and dataset in \autoref{fig:epsilon_sensitivity}.

An adversary's ASR will increase with respect to distortion budget because the adversary is given more space to maneuver. Thus, an ASR corresponding to $\epsilon = 0.0$ is a representation of one minus the victim model's accuracy on unperturbed data. Indeed, we observe that the trained RL Max Loss agent's performance increases dramatically with respect to $\epsilon$. Depending on the threat model and the adversary's distortion budget goal, we can observe the performance/budget trade-off when selecting $\epsilon$ to train the agent.

\noindent\textbf{Analyzing the effect of $c$.}
In our setup of the RL Min Norm attack, the agent is rewarded by $\Delta_{t+1}f + c \cdot  \Delta_{t+1}\delta$ at each timestep through $R_{\text{Min Norm}}$ and $P_{\text{Min Norm}}$. We run the RL Min Norm attack for 1200 PPO policy updates with $c\in\{10^{-3}, 10^{-2}, 10^{-1}\}$ and each tested on the training images from $\mathcal{D}$. We evaluate the policies trained from 3 random seeds and plot the average ASR (left axis) and $\ell_2$ distortion (right axis) with respect to $c$ on each victim model and dataset in \autoref{fig:c_sensitivity}.

The adversary's performance and distortion will decrease with respect to $c$ because the agent values decreasing the distortion over decreasing victim model confidence. Indeed, we observe this behavior as $c$ increases on log-scale. These results are consistent with traditional Min Norm attacks in AML~\citep{carlini_towards_2017} and agents trained under the RL Min Norm approach must consider this before or during training adversarial agents. Like our analysis on $\epsilon$ for RL Max Loss, an optimal value of $c$ depends on the adversary's goals and capabilities.

\subsection{Utility Analysis}\label{comparison}
Each adversarial agent trains on images from the training dataset $\mathcal{D}$. To investigate a broader landscape of attack capabilities in black-box settings, we ask: \textit{how well do trained adversarial agents generalize to unseen data, and how does it compare to traditional black-box methods?} We bifurcate the analysis of this question into two parts: (1) analyzing the adversarial samples crafted on an unseen dataset and (2) comparing the performance on unseen data of the RL attacks against known highly-performant black-box attacks on image classification. The first part will examine adversarial samples crafted after training on the training dataset $\mathcal{D}$ and testing dataset $\mathcal{D}^\prime$. The second part will compare the RL attacks against the black-box attack baselines over ASR, AQ, and average $\ell_2$-norm distortion. In \autoref{tab:attack_comparison}, we take RL Max Loss and RL Min Norm agents after 1 200 PPO policy updates training on dataset $\mathcal{D}$ in CIFAR-10 and SVHN tasks and evaluate the trained agents on the respective training and testing datasets $\mathcal{D}$ and $\mathcal{D}^\prime$. Additionally, we evaluate the baselines on the testing datasets $\mathcal{D}^\prime$ in both tasks. This is to represent a comparison across attack methods where all have no experience attacking the same data. We report the ASR, AQ, and average $\ell_2$-norm distortion ($\ell_2$) across all methods and datasets.

\noindent\textbf{Generalization.}
Here we evaluate whether the attack generalizes to unseen inputs, i.e., samples not used for training. Our experimental setup constructs disjoint sets $\mathcal{D}$ and $\mathcal{D}^\prime$ for training and testing, respectively. After training the attacks, we evaluate the trained policies on each sample from $\mathcal{D}$ and $\mathcal{D}^\prime$ on attack success, query count, and $\ell_2$-norm distortion. In \autoref{tab:attack_comparison}, we record the ASR, AQ, and $\ell_2$ distortion for each trained RL attack type against train and test datasets $\mathcal{D}$ and $\mathcal{D^\prime}$ from CIFAR-10 and SVHN.

We see that adversarial sample metrics are similar in distribution when comparing performance from training data $\mathcal{D}$ to testing data $\mathcal{D}^\prime$ in both CIFAR-10 and SVHN tasks across all victim models. This means the performance on the test dataset $\mathcal{D}^\prime$ is well within the distribution of the train dataset $\mathcal{D}$ performance. Thus, the policy learned by the agent generalizes to well to unseen inputs.

\noindent\textbf{Comparison.}
Given the RL attacks' ability to improve adversarial sample performance in black-box settings and generalize to new data, we compare adversarial agent generalization against known black-box attack baselines.We selected Square~\cite{andriushchenko_square_2020}, HopSkipJumpAttack (HSJA)~\cite{chen_hopskipjumpattack_2020}, and the Bandits~\cite{ilyas_prior_2019} attack, as these are prominent, state-of-the-art methods that represent different strategies for solving the black-box optimization problem (e.g., random search, decision-based, and prior-guided). To this end, we measure their ability to successfully craft adversarial samples on test dataset $\mathcal{D}^\prime$ from CIFAR-10 and SVHN by reporting ASR, AQ, and average $\ell_2$ distortion in the last three rows of \autoref{tab:attack_comparison}. Indeed, we see that the ASR and AQ of both RL Max Loss and RL Min Norm on test dataset $\mathcal{D}^\prime$ outperform the most competitive baseline in most tasks with as much or fewer $\ell_2$ distortion. Notably, achieving up to 17\% more attack success with 31\% fewer queries in the most extreme case (\texttt{VGG-16} on SVHN). Adversarial agent performance on new, unseen data is an artifact of why an adversary is likely to use an RL approach in a sustained attack setting where experience improves general attack performance.
\section{Discussion}\label{discussion}

\textbf{Sample Transferability.} A key area of black-box AML research is model transferability: the degree to which adversarial samples on one model are adversarial in another. In this work, we investigate an agentic approach to query-based black-box attacks, focused on whether a RL policy (the agent) can learn a robust strategy for generating attacks. Our results demonstrate that this is effective and generalizable to traditional image classification benchmarks and state-of-the-art image classification models. These findings suggest that adversarial agents can learn attack algorithms through RL. This RL framework recasts attack transferability as a more promising opportunity with policy adaptation. The extensive training of the RL policy on an initial victim model can be viewed as an effective pre-training step. We hypothesize that this pre-trained RL policy could then be adapted (or fine-tuned) using the same RL framework on a new victim model with greater attack success and query efficiency than a policy trained from scratch. This approach remains a logical and compelling direction for future work.

\noindent\textbf{MDP improvements.} The RL attacks are only as good as the action space and reward function that allow the agent to change the state of the input with more reward. The current formulation selects $N$ features of an input with each selected feature getting $\delta_i, i\in\{1,...,N\}$ with bounded distortion $|\delta_i|\leq\theta$ where $\theta\in(0,1)$ for the image domain. Distortion strategies come in many forms in AML, and alternative representations of the action space such as latent-space distortions could enhance the efficiency and attack success in RL. Further, including curriculum learning or adaptive rewards based on attack success could accelerate training. Future work should explore the combinations of different MDP components to develop more practical, real-world adversarial testing.

\noindent\textbf{Cross-domain applicability.} The study in this paper focuses on image classifiers, but the framework and analysis is not inherently tied to it. Images provide a natural domain to evaluate new attacks due to well-defined distortion metrics like the $\ell_2$-norm and rudimentary constraints that require features to be within $[0,255]$. Nevertheless, these strategies can be readily applied to other classification tasks such as malware and network intrusion detection. Investigating these domains not only studies the domain generalizability of the RL attacks proposed in this paper but also expands their impact on securing a range of machine learning applications.
\section{Conclusion}\label{conclusion}
This paper explores the capabilities of RL in black-box AML. We developed a novel MDP framework consisting of two agents, RL Max Loss and RL Min Norm, that emulate two classes of AML algorithms. The MDP setup enables the adversary to use RL to learn attack strategies that generate adversarial samples more successfully and with efficiency. Indeed, we find that: (1) the agents learn a policy that improves the attack success rate by up to 13.2\% and (2) the adversarial samples generated by the agents require up to 16.9\% fewer queries over training on benchmark image classification datasets. Further, we analyze the sensitivity of reward hyperparameters $\epsilon$ and $c$ for RL Max Loss and RL Min Norm to show their consistency with state-of-the-art AML algorithms. Last, we demonstrate the trained policy's ability to generalize crafting adversarial samples to inputs outside of the training dataset and show that in a comparison with state-of-the-art black-box attacks, the RL attacks become 17\% more successful at generating adversarial samples on unseen inputs post-training. These findings suggest that better black-box evasion attack strategies can be learned through RL, and that with the improvement of RL as a field results in stronger adversaries.

\section*{Acknowledgments}
The authors thank Kunyang Li, Yohan Beugin, and Nicholas Corrado as well as all reviewers for their helpful comments on previous iterations of the work.

\noindent\textbf{Funding acknowledgment.} This material is based upon work supported by the National Science Foundation under Grant No. CNS-2343611. Any opinions, findings, and conclusions or recommendations expressed in this material are those of the author(s) and do not necessarily reflect the views of the National Science Foundation.

{
    \small
    \bibliographystyle{ieeenat_fullname}
    \bibliography{refs}
}

\clearpage

\section*{Supplementary Material}\label{supplementary_material}

This supplementary material provides additional details on the experimental setup, including the selection of action hyperparameters and qualitative examples of generated adversarial samples.

\subsection*{Action Hyperparameters}\label{action_configs}
\begin{figure}[h]
    \centering
    \includegraphics[width=\linewidth]{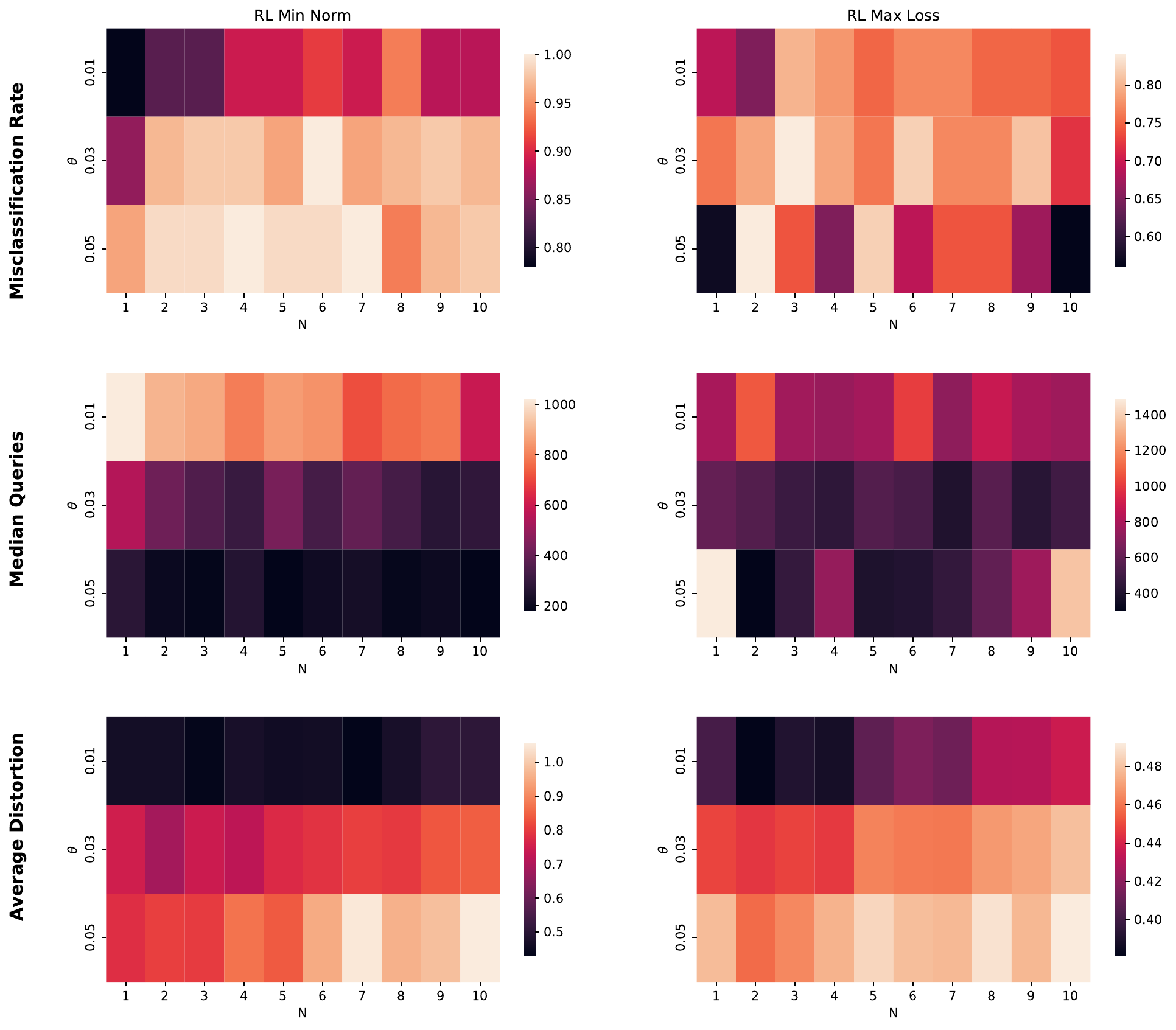}
    \caption{Misclassification Rate, Median Queries, and Average $\ell_2$-norm Distortion on adversarial examples post-training for different $(N,\theta)$ configurations with RL Min Norm and RL Max Loss attacks on the CIFAR-10 dataset.}
    \label{fig:ntheta_heatmaps}
\end{figure}

In Section 3.1, we define the action $a_t$ as a set of $N$ feature-perturbation pairs, $\{(i_1, \delta_1), ..., (i_N, \delta_N)\}$, where each perturbation $\delta_j$ has a maximum magnitude of $\theta$. The choice of $N$ (the number of features to perturb) and $\theta$ (the maximum magnitude of that perturbation) balances the trade-off between attack success and the complexity of the action space.

To select appropriate values, we conducted preliminary experiments, the results of which are shown in \autoref{fig:ntheta_heatmaps}. This figure plots the misclassification rate (ASR), median queries, and average $\ell_2$-norm distortion for various $(N, \theta)$ configurations for both RL Max Loss and RL Min Norm attacks on CIFAR-10. Based on these results, we fixed $N=5$ and $\theta=0.05$ for all main experiments in the paper, as this configuration offered a good balance between attack success and action complexity.

\subsection*{Hyperparameters}\label{hyperparams}

\begin{table}[t]
\centering
\setlength{\tabcolsep}{5pt} 
\begin{tabular}{@{}ll@{}} 
\toprule
\textbf{Parameter} & \textbf{Value} \\
\midrule
\multicolumn{2}{c}{\textbf{PPO Agent (\texttt{EfficientNet})}} \\
\cmidrule(lr){1-2}
Policy/Value Arch. & Linear(128, 64) \\
Optimizer & Adam \\
Learning Rate (LR) & 2.5e-3 \\
Discount Factor ($\gamma$) & 0.99 \\
GAE Lambda ($\lambda$) & 0.95 \\
Clip Range & 0.1 \\
\midrule
\multicolumn{2}{c}{\textbf{Victim Models (\texttt{ResNet50, VGG16})}} \\
\cmidrule(lr){1-2}
Optimizer & SGD \\
Momentum & 0.9 \\
Learning Rate (LR) & 0.001 \\
LR Scheduler & ReduceLROnPlateau \\
Weight Decay & 1e-4 \\
\midrule
\multicolumn{2}{c}{\textbf{Victim Model (\texttt{ViT\_B\_16})}} \\
\cmidrule(lr){1-2}
Optimizer & AdamW \\
Learning Rate (LR) & 0.001 \\
LR Scheduler & ReduceLROnPlateau \\
Weight Decay & 1e-4 \\
\bottomrule
\end{tabular}
\caption{Key hyperparameters for the PPO agent and the victim models. For the PPO agent, the learning rate and clip range are linearly annealed from their initial values to 0 over the course of training.}
\label{tab:hyperparams_condensed}
\end{table}

\autoref{tab:hyperparams_condensed} details the key hyperparameters used for training both the PPO agent and the victim models. The PPO agent's policy and value functions use an \texttt{EfficientNet} feature extractor. The victim models were fine-tuned from ImageNet-1K pre-trained weights.

\subsection*{RL Generated Adversarial Samples}\label{rl_adversarial_examples}
\begin{figure}[h]
    \centering
    \includegraphics[width=\linewidth]{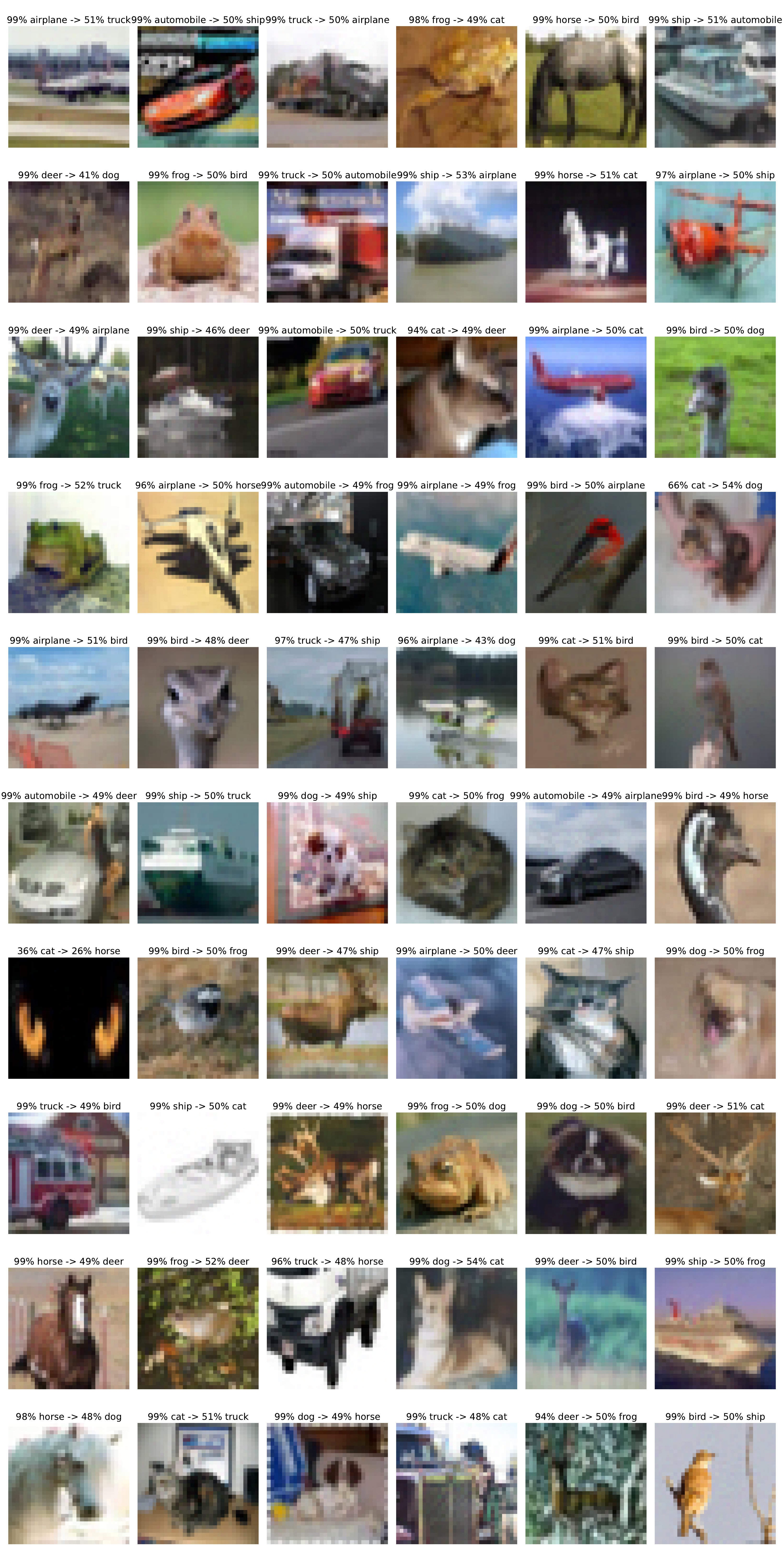}
    \caption{CIFAR-10 adversarial samples generated by black-box RL attacks. Each image contains the confidence on the original class and confidence on the incorrect class.}
    \label{fig:adversarial_examples}
\end{figure}

To provide a qualitative sense of the attacks, \autoref{fig:adversarial_examples} visualizes several adversarial samples generated by our trained RL agents on the CIFAR-10 dataset. Each example shows the original image, the resulting adversarial image, and the imperceptible perturbation (magnified for visibility). The labels demonstrate the agent's success: the victim model's confidence is shifted from the high-confidence original class to a high-confidence incorrect class. This aligns with the overall framework described in Figure 1, where the RL adversary iteratively queries the victim model to produce an adversarial example.

\end{document}